\documentstyle[epsf,epsfig,wrapfig,here,12pt]{article}    
\begin{document}           
\baselineskip=0.33333in
\begin{quote} \raggedleft TAUP 2832 - 2006
\end{quote}
\vglue 0.5in
\begin{center}{\bf The Yukawa Lagrangian Density\\
is Inconsistent with the Hamiltonian}
\end{center}
\begin{center}E. Comay$^*$
\end{center}

\begin{center}
School of Physics and Astronomy \\
Raymond and Beverly Sackler Faculty of Exact Sciences \\
Tel Aviv University \\
Tel Aviv 69978 \\
Israel
\end{center}
\vglue 0.5in
\vglue 0.5in
\noindent
PACS No: 03.65.Pm, 13.75.Cs
\vglue 0.2in
\noindent
Abstract:

It is proved that no Hamiltonian exists for
the real Klein-Gordon field used in the Yukawa interaction.
The experimental side supports this conclusion.

\newpage
About 70 years ago, the Yukawa interaction was proposed as a quantum
mechanical interpretation of the nuclear force
(see [1], p. 78). This interaction
is derived from the Lagrangian density of a system of a Dirac field
and a Klein-Gordon (KG) field (see [2], p. 79)
\begin{equation}
{\mathcal L}_Y = {\mathcal L}_D + {\mathcal L}_{KG} - g\phi \bar {\psi}\psi .
\label{eq:LYUKAWA}
\end{equation}
Here the first term on the right hand side
represents the Lagrangian density of a free Dirac field
(see [2], p. 43)
\begin{equation}
{\mathcal L}_D = \bar {\psi } (i\gamma ^\mu \partial _\mu - m)\psi
\label{eq:LDIRAC}
\end{equation}
and the second term represents the Lagrangian density of a free KG field
(see [2], p. 16)
\begin{equation}
{\mathcal L}_{KG} = \frac {1}{2} (g^{\mu \nu}\phi _{,\mu}\phi _{,\nu}
                        -m^2\phi ^2).
\label{eq:LKG}
\end{equation}
The last term of $(\!\!~\ref{eq:LYUKAWA})$
represents the interaction. Since the Hamiltonian is a
Hermitian operator, the KG function $\phi $ used here is real.

In this work, Greek indices run from 0 to 3 and Latin indices run
from 1 to 3. The Lorentz metric is diagonal and its entries are
(1,-1,-1,-1). Units where $\hbar = c = 1$ are used.
The symbol
$_{,\mu}$ denotes the partial differentiation with respect to $x^\mu $.

Difficulties concerning the KG Lagrangian density 
of a complex KG function have been pointed
out recently. Thus, it is proved that a KG particle cannot interact
with the electromagnetic fields: an application of the linear
interaction $j^\mu A_\mu $, where the KG 4-current $\j^\mu $ is independent
of the external 4-potential $A_\mu $, fails [3]; if the quadratic
expression $(p^\mu -eA^\mu)(p_\mu - eA_\mu)$ is used then the inner
product of the Hilbert space of the KG wave function $\phi $ is
destroyed. In addition to that, there is no
covariant differential operator
representing the Hamiltonian of a complex KG particle [4].

Another difficulty is the inconsistency of the 4-force derived from the
Yukawa potential
\begin{equation}
u(r) = -g^2e^{-mr}/r
\label{eq:YUKAWAPOTENTIAL}
\end{equation}
with the relativistic requirement where
the 4-acceleration must be orthogonal to the 4-velocity
\begin{equation}
a^\mu v_\mu = 0.
\label{eq:VA}
\end{equation}
This requirement is satisfied by the electromagnetic
interaction, where the Lorentz force is
\begin{equation}
ma^\mu = eF^{\mu \nu }v_\nu.
\label{eq:LORENTZF}
\end{equation}
Here the electromagnetic field tensor is antisymmetric
$F_{\mu \nu }= A_{\nu ,\mu } - A_{\mu ,\nu }$ and this property
satisfies $(\!\!~\ref{eq:VA})$
\begin{equation}
a^\mu v_\mu = \frac {e}{m}F^{\mu \nu }v_\nu v_\mu = 0.
\label{eq:VALORENTZ}
\end{equation}

On the other hand, the scalar function $\phi $ cannot yield an
antisymmetric tensor. Therefore, the force found in the classical limit
of the Yukawa interaction is inconsistent with special relativity.

The purpose of the present work is to prove that the Lagrangian density
$(\!\!~\ref{eq:LYUKAWA})$ of the real KG field $\phi $ is inconsistent with 
the fundamental quantum mechanical equation
\begin{equation}
i\frac {\partial \phi }{\partial t} = H\phi.
\label{eq:HPSI}
\end{equation}
This task extends the validity range of
the proof of [4] where the complex KG
field is discussed.

The Euler-Lagrange equations of a given Lagrangian density
are obtained from the following general
expression (see [2], p. 16)
\begin{equation}
\frac {\partial }{\partial x^\mu} \frac {\partial {\mathcal L}}
{\partial \frac {\partial \phi}{\partial x^\mu }} -
\frac {\partial {\mathcal L}}{\partial \phi} = 0.
\label{eq:FIELDEQ}
\end{equation}
Applying $(\!\!~\ref{eq:FIELDEQ})$  to the KG function $\phi $ of
$(\!\!~\ref{eq:LYUKAWA})$, one obtains an {\em inhomogeneous} KG equation

\begin{equation}
(\Box + m^2)\phi = g\bar {\psi}\psi.
\label{eq:KGINH}
\end{equation}

The following argument proves that the Euler-Lagrange equation
$(\!\!~\ref{eq:KGINH})$ obtained from the Yukawa Lagrangian density
$(\!\!~\ref{eq:LYUKAWA})$ is inconsistent with the existence of a
Hamiltonian. Indeed, the term $\partial ^2/\partial t^2$ of
$(\!\!~\ref{eq:KGINH})$ and the independence of its right hand side on
the KG wave function $\phi $, prove that it is a {\em second order
inhomogeneous} partial differential equation. On the other hand,
the Hamiltonian equation $(\!\!~\ref{eq:HPSI})$ is a {\em first
order homogeneous} equation. Now, assume that at a certain instant $t_0$,
a solution $\phi _0$ of $(\!\!~\ref{eq:HPSI})$ solves 
$(\!\!~\ref{eq:KGINH})$ too. Using the fact that
$(\!\!~\ref{eq:KGINH})$ is a second order differential equation, 
one finds that its first
derivative with respect to the time is a free parameter. This degree
of freedom proves that an infinite number of different solutions of
$(\!\!~\ref{eq:KGINH})$ agree with the {\em single} solution $\phi _0$ of
$(\!\!~\ref{eq:HPSI})$ at $t_0$. Thus, for $t>t_0$, just one solution of
$(\!\!~\ref{eq:KGINH})$ agrees with the solution of the Hamiltonian
$(\!\!~\ref{eq:HPSI})$ and all other solutions differ from it.

Moreover, if $\phi _0$ solves the {\em homogeneous} equation
$(\!\!~\ref{eq:HPSI})$, then $c\phi _0$, where $c$ is a constant, solves
it too. Therefore, at $t_0$,
an infinite number of solutions that solve the
Hamiltonian equation $(\!\!~\ref{eq:HPSI})$ correspond to every
solution of the Euler-Lagrange equation $(\!\!~\ref{eq:KGINH})$ obtained
from the Yukawa Lagrangian density.

Either of these results prove that the Yukawa Lagrangian density
$(\!\!~\ref{eq:LYUKAWA})$ is inconsistent with the existence of a
Hamiltonian. It is interesting to note that the Dirac Hamiltonian
agrees perfectly with the Euler-Lagrange equation obtained from
the Dirac Lagrangian density $(\!\!~\ref{eq:LDIRAC})$ (see
[4], p. 32).

Another problem of the Yukawa Lagrangian density $(\!\!~\ref{eq:LYUKAWA})$
is that its wave function $\phi $ is real. Hence, 
the real Yukawa function $\phi $
cannot be an energy eigenfunction (namely, an eigenfunction 
of the operator $i\partial/\partial t$), 
because an energy eigenfunction has a 
complex factor $e^{-i\omega t}$.
Therefore, the Yukawa particle cannot be an isolated free particle. This
result provides a proof showing 
that $\pi ^0$ is not a Yukawa particle. Indeed,
the lifetime of $\pi ^0$  is about $10^{-16}$ 
seconds (see [5], p. 500). Thus, having a
relativistic velocity, its path is more than $10^7$ fermi. This length
is much larger than the nucleon's radius which is about 1.2 fermi. Hence,
$\pi ^0$ is a free particle for the most of its lifetime, contrary
to the above mentioned restriction on a Yukawa particle.

\begin{figure}[t]
\vspace*{3ex}
\begin{center}
\rotatebox{0} {\includegraphics*[height=7cm]{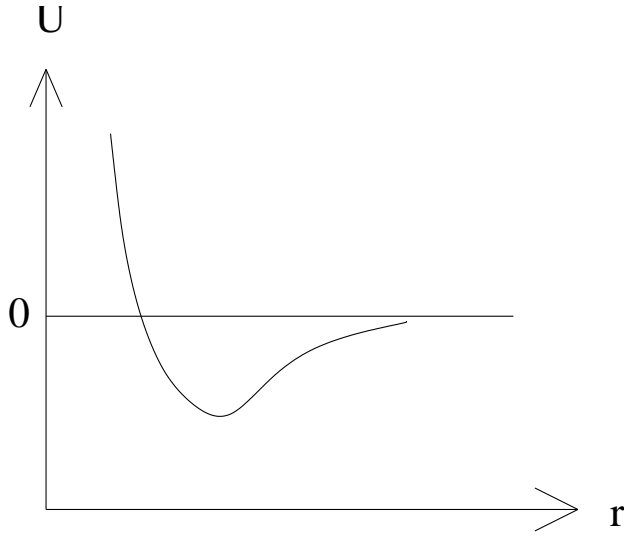}}
\end{center}
\begin{quotation}
\caption{
A qualitative description of the nucleon-nucleon phenomenological potential
as a function of the distance between the nucleons' centers.}


\end{quotation}
\vspace*{-1ex}
\end{figure}

Turning to the experimental side, it is not surprising to find that Nature
is unkind to the Yukawa theory. Thus,
the Klein-Gordon field function $\phi (x^\mu)$ depends on
a {\em single} set of space-time coordinates.
Hence, like the Dirac field $\psi (x^\mu)$, it describes a
structureless pointlike particle.
Now, unlike Dirac particles (electrons, muons, quarks etc.),
the existence of pointlike
KG particles has not been established. In particular, it is now
recognized that $\pi $ mesons, which are regarded as the primary
example of a KG particle, are made of a quarks 
and an antiquark. Hence, $\pi $ mesons are not pointlike
particles. Experimental data confirms this conclusion (see [5], p. 499).

The actual nuclear potential is inconsistent with the Yukawa
formula $(\!\!~\ref{eq:YUKAWAPOTENTIAL})$. Indeed, the nuclear
potential is characterized by a hard (repulsive) core and at its
outer side there is a rapidly decreasing attractive force. Its general
form is described in fig. 1 (see [1], p. 97).


Thus, the figure proves that the actual
nuclear potential and its derivative with respect to $r$
change sign. This is certainly inconsistent with the Yukawa formula
$(\!\!~\ref{eq:YUKAWAPOTENTIAL})$. Indeed, neither the Yukawa
potential nor its derivative change sign.

These arguments prove that the experimental side 
and the theoretical analysis carried out above, do not support the validity
of the Yukawa theory.


\newpage
References:
\begin{itemize}

\item[{*}] Email: elic@tauphy.tau.ac.il \\
           Internet site: http://www-nuclear.tau.ac.il/$\sim $elic
\item[{[1]}] S. S. M. Wong, {\em Introductory Nuclear Physics} (Wiley, 
New York, 1998). 2nd edition.
\item[{[2]}] M. E. Peskin and D. V. Schroeder, {\em An Introduction to 
Quantum Field Theory} (Addison-Wesley, Reading, Mass., 1995).
\item[{[3]}] E. Comay, Apeiron, {\bf 11}, No. 3, 1 (2004).
\item[{[4]}] E. Comay, Apeiron {\bf 12}, no. 1, 27 (2005).
\item[{[5]}] S. Eidelman et al. (Particle Data Group), Phys. Lett. {\bf B592}, 1 (2004).

\end{itemize}

\end{document}